\documentclass[aps,onecolumn,superscriptaddress,showpacs]{revtex4}
\usepackage{times,epsfig,amssymb,amsfonts,amsmath}
\usepackage{bm}

\begin{document}

\title{Theory of cavity ring-up spectroscopy}
\author{Ming-Yong Ye}
\email{myye@fjnu.edu.cn}
\affiliation{Fujian Provincial Key Laboratory of Quantum Manipulation and New Energy Materials, College of Physics and Energy, Fujian Normal University, Fuzhou 350117, China}
\affiliation{Fujian Provincial Collaborative Innovation Center for Optoelectronic Semiconductors and Efficient Devices, Xiamen 361005, China}
\affiliation{Key Laboratory of Quantum Information, University of Science and Technology of China, Chinese Academy of Sciences, Hefei 230026, China}
\author{Xiu-Min Lin }
\affiliation{Fujian Provincial Key Laboratory of Quantum Manipulation and New Energy Materials, College of Physics and Energy, Fujian Normal University, Fuzhou 350117, China}
\affiliation{Fujian Provincial Collaborative Innovation Center for Optoelectronic Semiconductors and Efficient Devices, Xiamen 361005, China}
\begin{abstract}
Cavity ring-up spectroscopy (CRUS) provides an advanced technique to sense
ultrafast phenomena, but there is no thorough discussion on its theory.
Here we give a detailed theoretical analysis of CRUS with and without modal coupling,
and present exact analytical expressions for the normalized transmission, which are very simple under certain reasonable conditions.
Our results provide a solid theoretical basis for the applications of CRUS.
\end{abstract}
\maketitle

\section{Introduction}
Whispering-gallery-mode (WGM) microresonators confine light through internal total reflection and can support optical modes with high quality  factors and small mode volumes \cite{Righini2011}. These features make them an important platform for studying the interaction between light and matter \cite{keppenberg2007,shendong2016,wang2017,wu2014,park2006}.
When a continuous laser is utilized to sweep the optical modes of a WGM microresonator via a fiber taper at very slow speed,
the steady-state transmission spectrum of the system can be obtained \cite{knight1997phase}.
By monitoring the change of the steady-state spectrum, such as the mode shift \cite{vollmer2008}, broadening \cite{shao2013} and splitting \cite{zhu2009}, highly sensitive sensing can be achieved  \cite{kimvollmer,zhi2017}.
However, these methods can only sense the environmental change on a timescale of milliseconds due to the slow sweeping speed of the laser.

Recently cavity ring-up spectroscopy (CRUS) is proposed \cite{rosenblum2015}. Experiments with WGM microresonators show that CRUS can be measured within tens of nanoseconds
and thus it provides a way to achieve  ultrafast sensing  \cite{rosenblum2015}.
CRUS can be understood through an abrupt turn on of a far-detuned monochromatic laser that is connected to the input of the fiber taper.
Although the laser frequency is far detuned from the mode frequency of the microresonator, some light of the mode frequency can still be coupled into the microresonator due to
the  broadening of the laser frequency, which is caused by the sharp rise of the laser intensity.
The light into the microresonator will leak back to the fiber taper and
then interfere with the directly transmitted light, which can lead to an oscillation of the light intensity in the output of the fiber taper.
As a demonstration of its ability in sensing ultrafast phenomena, CRUS has shown its application in monitoring the optomechanical vibrations of the WGM microresonator \cite{rosenblum2015}.
It is believed that CRUS can be used in the study of ultrafast phenomena such as protein folding, light harvesting and cavity quantum electrodynamics \cite{rosenblum2015}.
However, there is no detailed theory of CRUS that may limit its future application \cite{yang2016}.
It is the purpose of this paper to provide a theory of CRUS.
\begin{figure}[htbp]
\centering\includegraphics[width=9 cm]{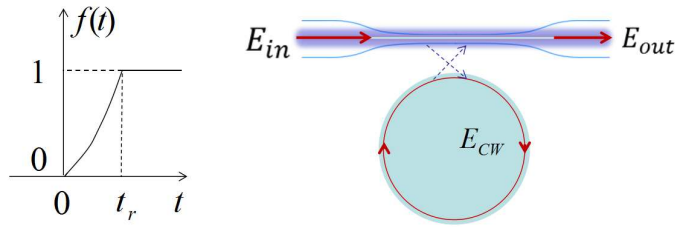}
\caption{Schematic illustration of a WGM microresonator coupled to a fiber taper (right) and a step-like modulation function $f(t)$ (left).}
\end{figure}

\section{Theory of CRUS}
The idea of the CRUS is illustrated in Fig. 1, where a fiber taper is used to couple light into and out of a WGM microresonator.
The input field in the fiber taper is denoted by $E_{in}$, which will excite the clockwise (CW) mode in the microresonator
denoted by $E_{cw}$. The evolution of $E_{cw}$ satisfies the equation \cite{Righini2011}
\begin{equation}
\frac{dE_{cw}(t)}{dt}=(-jw_c-\kappa)E_{cw}(t)+ \sqrt{2\kappa_e}E_{in}(t),  \label{q1}
\end{equation}
where $w_c$ is the angular frequency of the CW mode, $\kappa$ is the total loss rate, and $\kappa_e \,(\kappa_e<\kappa)$ is the loss rate associated with the fiber taper.
In CRUS, a monochromatic laser with a special amplitude modulation is coupled into the fiber taper and
the input field can be written as
\begin{equation}
  E_{in}(t)=f(t)* s e^{-jw_l t}, \label{q2}
\end{equation}
where $s$ is the amplitude of the monochromatic laser, $w_l$ is the angular frequency of the monochromatic laser and $f(t)$ is a step-like modulation function as shown in Fig. 1.
The value of the function $f(t)$ is $0$ when $t\leq0$, changes from $0$ to $1$ in a short rise time $t_r$, and keeps the value $1$ when $t\geq t_r$.
The input in Eq. (\ref{q2}) describes the abrupt turn on of a monochromatic laser at $t=0$.

It is natural to assume that $E_{cw}(0)=0$. Based on the method of variation of constants, an integral expression for $E_{cw}(t)$ can be obtained,
\begin{equation}
E_{cw}(t)=e^{(-jw_c-\kappa)t}\int_0^t e^{(jw_c+\kappa)t'}\sqrt{2\kappa_e}E_{in}(t')dt'. \label{int}
\end{equation}
The correctness of Eq. (\ref{int}) can be verified by the facts that it satisfies the initial condition $E_{cw}(0)=0$ and the differential equation in Eq. (\ref{q1}).
Substitute Eq. (\ref{q2}) into Eq. (\ref{int}), there is
\begin{equation}
E_{cw}(t)=s\sqrt{2\kappa_e} e^{(-jw_c-\kappa)t} \int_0^t e^{(-j\delta+\kappa)t'}f(t')dt',
\end{equation}
where $\delta=w_l-w_c$ is the detuning of the angular frequency of the laser relative to the angular frequency of the CW mode.
Using the method of integration by parts and $f(0)=0$, we can obtain
\begin{equation}
E_{cw}(t)=\frac{s\sqrt{2\kappa_e}}{-j\delta+\kappa} e^{(-jw_c-\kappa)t} \{e^{(-j\delta+\kappa)t}f(t)-\int_0^t e^{(-j\delta+\kappa)t'}\frac{df(t')}{dt'}dt' \}.
\end{equation}
The interesting time interval in CRUS is $t\geq t_r$ where there is $f(t)=1$ and $df(t)/dt=0$. Define
\begin{equation}
\alpha=\int_{0}^{t_r} e^{(-j\delta+\kappa)(t'-t_r)}\frac{df(t')}{dt'} dt'. \label{alpha}
\end{equation}
There is
\begin{eqnarray}
E_{cw}(t)&=&\frac{s\sqrt{2\kappa_e}}{-j\delta+\kappa} e^{(-jw_c-\kappa)t} \{e^{(-j\delta+\kappa)t}-\alpha e^{(-j\delta+\kappa) t_r} \},\,t\geq t_r, \\
        &=&\frac{\sqrt{2\kappa_e}}{-j\delta+\kappa}  \{1-\alpha e^{(j\delta-\kappa) (t-t_r)} \}E_{in}(t),\,t\geq t_r,  \label{qcw}
\end{eqnarray}
which has two terms, one with the angular frequency $w_l$ of the laser and the other with the angular frequency $w_c$ of the CW mode.
The parameter $\alpha$
describes
the effect of $f(t)$ on the field $E_{cw}(t)$.
A  detailed discussion on the parameter $\alpha$ will be given in section 4,
where we will show that there is $|\alpha|<1$, and $\alpha \approx 1$ can be achieved if the rise time $t_r$ is short enough.

The output field $E_{out}$ in the fiber taper is related to the input field and the field in the microresonator through the expression \cite{Righini2011}
\begin{equation}
E_{out}(t)=-E_{in}(t)+\sqrt{2\kappa_e}E_{cw}(t). \label{out}
\end{equation}
Substitute Eq. (\ref{qcw}) into the above, we can get
\begin{equation}
E_{out}(t)=\{Ae^{j \theta_A}-Be^{j \theta_B} e^{(j\delta-\kappa) (t-t_r)}\}E_{in}(t),\,t\geq t_r,
\end{equation}
where the complex numbers $Ae^{j \theta_A}$ and $Be^{j \theta_B}$ are written in the polar form, and there are
\begin{equation}
Ae^{j \theta_A}=-1+\frac{2\kappa_e}{-j\delta+\kappa},  \:
Be^{j \theta_B}=\alpha \frac{2\kappa_e}{-j\delta+\kappa}.
\end{equation}
The normalized transmission $T(t)=\left |E_{out}(t)/E_{in}(t)\right |^2 $ will be
\begin{equation}
T(t)=A^2+B^2e^{-2\kappa (t-t_r)}-2ABe^{-\kappa (t-t_r)}cos\{\delta (t-t_r)+\theta_B-\theta_A\}, \, t\geq t_r, \label{main}
\end{equation}
which describes the content of CRUS.
The third term in the above shows that  $T(t)$ will oscillate with the period $T_p=2\pi/|\delta|$, the oscillation lifetime $\tau=1/\kappa$ and the initial
oscillation amplitude $A_p=\left | 2AB \right |$. Within the lifetime $\tau$, the number of the oscillation will be $N_p=\tau /T_p=|\delta| /(2\pi \kappa)$,
which indicates that $|\delta| \gg \kappa$ is necessary even when we only want to observe a small number of oscillations within the lifetime.

Suppose the condition $|\delta| \gg \kappa$ is satisfied,
then $Ae^{j \theta_A}  \approx -1$, $Be^{j \theta_B}  \approx  2\kappa_e \alpha j/ \delta$ and
\begin{equation}
E_{out}(t)\approx \{-1- \frac{2\kappa_e \alpha}{ \delta} j e^{(j\delta-\kappa) (t-t_r)}\}E_{in}(t),\,t\geq t_r,
\end{equation}
which will lead to
\begin{equation}
T(t)\approx 1- \frac{4\kappa_e}{\delta} |\alpha|  e^{-\kappa (t-t_r)}sin\{\delta (t-t_r)+\theta_{\alpha}\},\, t\geq t_r, \label{tt}
\end{equation}
where $\theta_{\alpha}$ is the phase of $\alpha$ and the term  proportional to $(\kappa_e/\delta)^2$ is omitted. Since  the initial oscillation amplitude of $T(t)$ is $A_p=4\kappa_e |\alpha| /|\delta|$,
it is better to experimentally observe CRUS in the over coupling condition with an optical mode of ultrahigh intrinsic quality factor so that $\kappa_e$ can be large enough to get a relative large $A_p$.

\section{Theory of CRUS with modal coupling}
A WGM microresonator can also support a counterclockwise (CCW) mode that is degenerate to the CW mode.
Now we consider the case where there is a coupling
between the degenerate CW and CCW modes  \cite{kippenberg2002}. The CCW mode is denoted by $E_{ccw}$. Instead of the Eq. (\ref{q1}), the dynamical equations are  \cite{kippenberg2002}
\begin{eqnarray}
\frac{dE_{cw}(t)}{dt}&=&(-jw_c-\kappa)E_{cw}(t)+j\beta E_{ccw}(t)+ \sqrt{2\kappa_e}E_{in}(t),  \label{qq} \\
\frac{dE_{ccw}(t)}{dt}&=&(-jw_c-\kappa)E_{ccw}(t)+j\beta E_{cw}(t),
\end{eqnarray}
where $\beta$ is the coupling strength between the CW and CCW modes.
Define
\begin{equation}
E_{+}(t)=E_{cw}(t)+ E_{ccw}(t), \:
E_{-}(t)=E_{cw}(t)- E_{ccw}(t).
\end{equation}
It can be obtained that
 \begin{eqnarray}
\frac{dE_{+}(t)}{dt}&=&\{-j(w_c-\beta)-\kappa \}E_{+}(t)+ \sqrt{2k_e}E_{in}(t),  \label{eplus} \\
\frac{dE_{-}(t)}{dt}&=&\{-j(w_c+\beta)-\kappa \} E_{-}(t)+ \sqrt{2k_e}E_{in}(t),\label{eminus}
\end{eqnarray}
which are decoupled equations similar to Eq. (\ref{q1}).
Therefore similar to Eq. (\ref{qcw}) there are
 \begin{eqnarray}
E_{+}(t)&=&\frac{\sqrt{2\kappa_e}}{-j\delta_{+}+\kappa}  \{1-\alpha_{+} e^{(j\delta_{+}-\kappa) (t-t_r) } \}E_{in}(t),\, t\geq t_r, \\
E_{-}(t)&=&\frac{\sqrt{2\kappa_e}}{-j\delta_{-}+\kappa}  \{1-\alpha_{-} e^{(j\delta_{-}-\kappa) (t-t_r) } \}E_{in}(t),\, t\geq t_r,
 \end{eqnarray}
where $\delta_{+}=\delta+\beta$, $\delta_{-}=\delta-\beta$ and
\begin{equation}
\alpha_{+}=\int_{0}^{t_r} e^{(-j\delta_{+}+\kappa)(t'-t_r)}\frac{df(t')}{dt'} dt',  \:
\alpha_{-}=\int_{0}^{t_r} e^{(-j\delta_{-}+\kappa)(t'-t_r)}\frac{df(t')}{dt'} dt'.
\end{equation}
Note that $E_{cw}(t)=\{E_{+}(t)+E_{-}(t)\}/2$.
According to Eq. (\ref{out}), the exact expression for the normalized transmission $T(t)=|E_{out}(t)/E_{in}(t)|^2$ is
 \begin{equation}
T(t)=\left |C-\frac{\kappa_e}{-j\delta_{+}+\kappa}\alpha_{+} e^{(j\delta_{+}-\kappa) (t-t_r) }
-\frac{\kappa_e}{-j\delta_{-}+\kappa}\alpha_{-} e^{(j\delta_{-}-\kappa) (t-t_r) } \right |^2,  \, t\geq t_r, \label{ex}
 \end{equation}
where
\begin{equation}
C=-1+\frac{\kappa_e}{-j\delta_{+}+\kappa}+\frac{\kappa_e}{-j\delta_{-}+\kappa}.
\end{equation}
In the following, with some approximations we will present a simple expression of $T(t)$.

Assume $\beta t_r\ll 1$, $\kappa\ll |\delta|$ and $\beta\ll |\delta|$, there will be $\alpha_{+}\approx \alpha_{-} \approx \alpha=|\alpha|e^{j\theta_{\alpha}}$ and
\begin{equation}
\frac{\kappa_e}{-j\delta_{+}+\kappa}\approx \frac{\kappa_e}{-j\delta_{-}+\kappa} \approx \frac{\kappa_e}{\delta}j.
\end{equation}
Then we can obtain
\begin{eqnarray}
T(t) & \approx & \left |-1-\frac{\kappa_e \alpha}{\delta}j  e^{(j\delta_{+}-\kappa) (t-t_r) }
-\frac{\kappa_e \alpha}{\delta}j e^{(j\delta_{-}-\kappa) (t-t_r) } \right |^2,  \, t\geq t_r, \\
& \approx & 1-\frac{4 \kappa_e}{\delta}|\alpha|e^{-\kappa (t-t_r) }cos[\beta(t-t_r)]sin[\delta(t-t_r)+\theta_{\alpha}],\, t\geq t_r, \label{mc}
\end{eqnarray}
where the terms proportional to $(\kappa_e/\delta)^2$ are omitted. As $\beta\ll |\delta|$,
it can be seen that $T(t)$ oscillates with the amplitude
$|(4 \kappa_e/\delta) \alpha e^{-\kappa (t-t_r) }cos[\beta(t-t_r)]|$ and the period $2\pi/|\delta|$.
The amplitude decays with the rate $\kappa$ and oscillates slowly with a period $\pi/\beta$. To show the validity of the approximate $T(t)$ in Eq. (\ref{mc}), it is compared with
the exact $T(t)$ in Eq. (\ref{ex}) in Fig. 2.
It can be seen that the approximate $T(t)$ agrees well with the exact one.
\begin{figure}[htbp]
\centering\includegraphics[width=10cm]{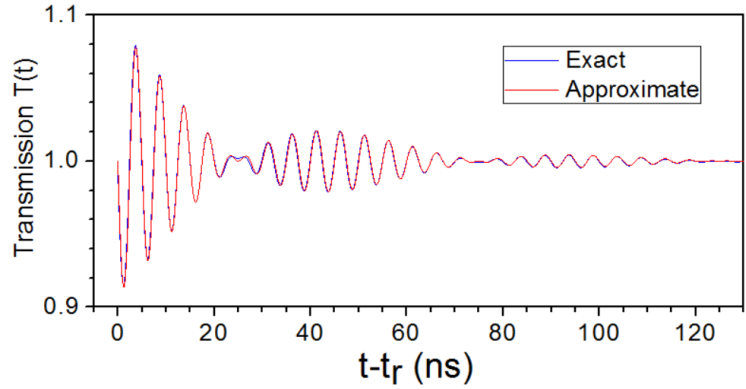}
\caption{Comparison between the exact $T(t)$ in Eq. (\ref{ex}) and the approximate $T(t)$ in Eq. (\ref{mc}) with $\kappa_e/2\pi=4.5$ MHz, $\kappa/2\pi=5$ MHz, $\beta/2\pi=10$ MHz and
$\delta/2\pi=200$ MHz. For exact $T(t)$, $\alpha_{+}$ and $\alpha_{-}$ are calculated from $f(t)=t/t_r$ ($0\leq t \leq t_r$) with $t_r=0.1$ ns.
For approximate $T(t)$, $\alpha=1$ is used (see the section 4).}
\end{figure}

\section{Influence of the modulation function}
The modulation function $f(t)$ affects the normalized transmission $T(t)$ through the parameter $\alpha$.
We first note that there is $|\alpha|<1$, because from its definition in Eq. (\ref{alpha}),
\begin{equation}
\left | \alpha \right |<\int_{0}^{t_r}\left | e^{(-j\delta+\kappa)(t'-t_r)} \frac{df(t')}{dt'}\right | dt'<\int_{0}^{t_r} \frac{df(t')}{dt'} dt'=1,
\end{equation}
where $df(t)/dt\geq 0$ is assumed, which is reasonable in CRUS. If the rise time $t_r$ is so short that $|\delta|t_r \ll 1$ and $\kappa t_r \ll 1$,
then $\alpha \approx 1$ because from Eq. (\ref{alpha})
\begin{equation}
\alpha  \approx \int_{0}^{t_r} \frac{df(t')}{dt'} dt'=f(t_r)-f(0)=1,
\end{equation}
where only the first term of the Taylor series of the exponential function in Eq. (\ref{alpha}) is kept.
The result means that CRUS will not depend on the detailed shape of the modulation function $f(t)$ if the rise time $t_r$ is short enough.
A correction to $\alpha \approx 1$ can be obtained if the first two terms of  the Taylor series of  the exponential function in Eq. (\ref{alpha}) are kept, i.e.,
\begin{equation}
\alpha  \approx \int_{0}^{t_r}\{1+(-j\delta+\kappa)(t'-t_r)\} \frac{df(t')}{dt'} dt=1-\gamma (-j\delta+\kappa)t_r.
\end{equation}
The parameter $\gamma$ is defined as
\begin{equation}
\gamma  =-\frac{1}{t_r} \int_{0}^{t_r}(t'-t_r) \frac{df(t')}{dt'} dt'=\frac{1}{t_r}\int_{0}^{t_r}f(t')dt',
\end{equation}
where the method of the integration by parts is used. It can be seen that there is $0<\gamma<1$.

For several special modulation function $f(t)$, analytical expressions of $\alpha$ can be calculated. In the case $f(t)=t/t_r$ ($0\leq t \leq t_r$),
from the definition of $\alpha$ in Eq. (\ref{alpha}) there is
\begin{equation}
\alpha=\int_{0}^{t_r} e^{(-j\delta+\kappa)(t'-t_r)}\frac{1}{t_r} dt'=\frac{1}{(-j\delta+\kappa)t_r}\{1-e^{(j\delta-\kappa)t_r}\}. \label{ex1}
\end{equation}
In the case $f(t)=sin(\frac{\pi t}{2 t_r})$ ($0\leq t \leq t_r$), there is
\begin{eqnarray}
\alpha&=&\frac{\pi }{2 t_r} \int_{0}^{t_r} e^{(-j\delta+\kappa)(t'-t_r)}cos(\frac{\pi t'}{2 t_r}) dt', \\
&=&\frac{\pi }{4}\frac{1}{(-j\delta'+\kappa)t_r}\{j-e^{(j\delta-\kappa)t_r}\}-\frac{\pi }{4}\frac{1}{(-j\delta''+\kappa)t_r}\{j+e^{(j\delta-\kappa)t_r}\},
\end{eqnarray}
where $\delta'=\delta-\pi /(2 t_r)$ and $\delta''=\delta+\pi /(2 t_r)$. These two exact expressions demonstrate that $\alpha \approx 1$
if the rise time $t_r$ is short enough.

We finally note that the discussions in the above two sections will not change when
$f(t)$ ($0\leq t \leq t_r$) is a complex function with $f(t_r)=1$ (this requirement will not make $f(t)$ loss of generality).
And we will still have $\alpha \approx 1$ if the rise time $t_r$ is short enough.

\section{Conclusion}
We have obtained exact and approximate expressions for the normalized transmission of CRUS.
The results show clearly how the parameters, such as the detuning, modal coupling, total loss rate, loss rate associated with the fiber taper and
modulation function, affect the normalized transmission.
The work will promote the application of CRUS.

\section*{Funding}
National Natural Science Foundation of China (Grant Nos. 11674059, 61275215); Fujian Provincial College Funds for Distinguished Young Scientists (Grant No. JA14070); Natural Science Foundation of Fujian Province (Grant Nos. 2016J01008, 2016J01009); Open Project of Key Laboratory of Quantum Information (Chinese Academy of Sciences) under Grant No. KQI201601.

\end{document}